\documentclass[aps,prl,twocolumn,superscriptaddress,showpacs]{revtex4}
\usepackage{graphicx}
\usepackage{natbib}
\usepackage{color}
\usepackage{amsmath}
\usepackage{amssymb}
\usepackage{array}

\begin{document}

%\preprint{}

\title{Pair Correlations, Short Range Order and Dispersive Excitations in the Quasi-Kagome Quantum Magnet Volborthite}


\author{G. J. Nilsen}
\email[Email address:~]{gorannilsen@gmail.com}
\affiliation{Laboratory for Quantum Magnetism, ICMP, Ecole Polytchnique F\'{e}d\'{e}rale de Lausanne (EPFL), Switzerland}
\affiliation{Department of Chemistry and Biochemistry, Universit\"{a}t Bern, Freiestrasse 3, CH-3012 Bern, Switzerland}
\affiliation{School of Chemistry, University of Edinburgh, West Mains Road, Edinburgh, EH9 3JJ, United Kingdom}

\author{F. C. Coomer}
\altaffiliation[Current Adress:~]{Dep. of Pure and Applied Chemistry, University of Strathclyde, Glasgow, G1 1XL UK}
\affiliation{School of Chemistry, University of Edinburgh, West Mains Road, Edinburgh, EH9 3JJ, United Kingdom}

\author{M. A. de Vries}
\affiliation{Department of Physics and Astronomy, University of Leeds, Leeds, LS2 9JT United Kingdom}

\author{J. R. Stewart}
\affiliation{ISIS facility, Rutherford Appleton Laboratory, STFC, Chilton, Didcot, OX11 0DE United Kingdom}

\author{P. P. Deen}
\affiliation{European Spallation Source, ESS AB, P. O. Box 176, SE-22100 Lund, Sweden}

\author{A. Harrison}
\affiliation{School of Chemistry, University of Edinburgh, West Mains Road, Edinburgh, EH9 3JJ, United Kingdom}
\affiliation{Institut Laue-Langevin, 6 rue Jules Horowitz, F-38042 Grenoble, France}

\author{H. M. R\o{}nnow}
\affiliation{Laboratory for Quantum Magnetism, ICMP, Ecole Polytchnique F\'{e}d\'{e}rale de Lausanne (EPFL), Switzerland}


\date{\today}
\begin{abstract}
We present spatial and dynamic information on the 
$s=1/2$ distorted kagome antiferromagnet volborthite, Cu$_3$V$_2$O$_7$(OD)$_2 \cdot 2$D$_2$O, obtained by polarized and inelastic neutron scattering. The instantaneous structure factor, $S(Q)$, is dominated by nearest neighbor pair correlations, with short range order at wave vectors $Q_1=0.65(3)$~\AA$^{-1}$ and $Q_2=1.15(5)$~\AA$^{-1}$ emerging below $5$~K. The excitation spectrum, $S(Q,\omega)$, reveals two steep branches dispersing from $Q_1$ and $Q_2$, and a flat mode at $\omega_f=5.0(2)$~meV. The results allow us to identify the cross-over at $T^\ast\sim1$~K  in $^{51}$V NMR and specific heat measurements as the build-up of correlations at $Q_1$. We compare our data to theoretical models proposed for volborthite, and demonstrate that the excitation spectrum can be explained by spin-wave-like excitations with anisotropic exchange parameters, as also suggested by recent local density calculations. 
\end{abstract}

\maketitle

The quantum kagome Heisenberg antiferromagnet (QKHAF) is among the most coveted targets in the quest for experimental realizations of quantum spin liquid ground states. The two most prominent physical realizations of the QKHAF studied to date are the naturally occuring minerals herbertsmithite, Cu$_3$Zn(OH)$_6$Cl$_2$ \cite{Shores2005} and volborthite, Cu$_3$V$_2$O$_7$(OH)$_2 \cdot 2$H$_2$O \cite{Hiroi2001}. The physics of herbertsmithite is arguably influenced by depletion of the kagome lattice caused by antisite mixing, possibly resulting in a valence bond glass state \cite{Vries2009,Singh2010}. In volborthite, the kagome planes are slightly distorted, but the lattice coverage is essentially complete. Since both systems deviate from the pure QKHAF model, focus has shifted to the intriguing question of which states arise when the QKHAF is perturbed. Remarkably, a multitude of different states have been proposed theoretically, depending on the nature of the perturbation. These range from ordered states \cite{Yavorskii2007,Wang2007,Cepas2008,Rousochatzakis2009,Messio2011} to the aforementioned valence bond glass. Experimentally, the ground state of volborthite remains enigmatic, despite over $9$ years of intensive study  \cite{Hiroi2001,Bert2005,Yoshida2009,Yoshida2009a,Yamashita2010}. Here we present a neutron scattering investigation of volborthite, employing both $xyz$ polarised and inelastic time of flight techniques.

\begin{figure}[h]
\includegraphics[width=0.9\linewidth]{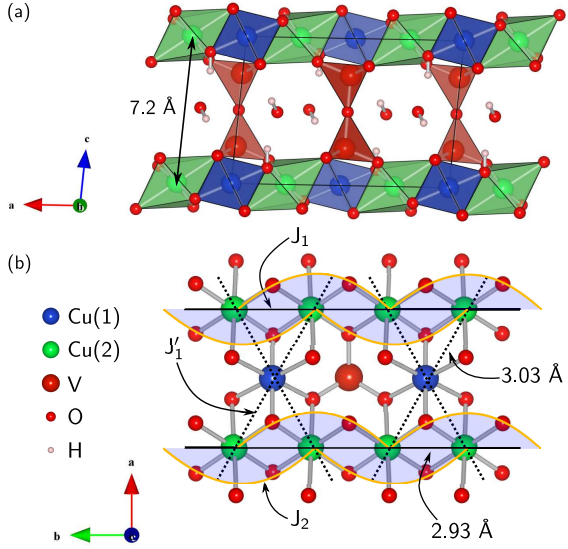}
\caption{Structure of volborthite. (a) Kagome planes of Cu(1) and Cu(2) octahedra (respectively blue and green) are separated by V$_2$O$_7$ columns (red). (b) Local environments of Cu$^{2+}$ in a kagome plane. Solid and dashed black lines indicate the two nearest neighbour exchanges, $J_1$ and $J_1'$. The next-nearest neighbour exchange $J_2$ along the $b$-direction is shown in orange.}
\label{fig:structure}
\end{figure}

Volborthite contains distorted kagome planes of edge sharing Cu$^{2+}$ octahedra, well separated ($\sim 7.2$~\AA) by pyrovanadate columns (V$_2$O$_7$),  Fig. \ref{fig:structure}(a). There are two crystallographically distinct Cu$^{2+}$ ions. Cu$(2)$ form chains along the $b$ directions whileCu$(1)$ populates the interchain sites. Locally, Cu$(1)$ and Cu$(2)$ reside in tetragonally and axially distorted octahedra, respectively, suggesting the $d_{\mathsf{3z^2-r^2}}$ orbital being singly occupied on the Cu$(1)$ site, with $d_{\mathsf{x^2-y^2}}$ the magnetically active orbital for Cu$(2)$. As a consequence, there are two different nearest neighbour exchange pathways, $J_1$ ($r_{\mathsf{Cu(2)-Cu(2)}}=2.93$~\AA) and $J_1'$ ($r_{\mathsf{Cu(1)-Cu(2)}}=3.03$~\AA). $J_1$ links Cu$(2)$ ions along the $b$ axis. $J_1'$ connects Cu$(1)$ and Cu$(2)$ ions. Furthermore, the edge sharing of the Cu$(2)$ octahedra along the $b$ direction imply the possibility of a strong next-nearest neighbour exchange, $J_2$.

Despite a large estimated average coupling $J_{\mathsf{avg}}=(2J_1'+J_1)/3=84$~K, magnetic susceptibility and specific heat measurements on volborthite show no signs of long range order down to $1.8$~K \cite{Hiroi2001}. At yet lower temperatures, muon spin rotation ($\mu$SR) and $^{51}$V NMR \cite{Bert2005,Yoshida2009,Yoshida2009a} indicate slowing down of fluctuations at $T^\ast=1$~K, but with dynamics persisting to $20$~mK. The state below $T^\ast$ was interpreted as either incommensurate or short range correlated. Consistent with this, low temperature specific heat studies indicate a high density of low energy modes below $T^\ast$ \cite{Yamashita2010}. 

These results, however, have provided little insight to the nature of the magnetic correlations and excitations. Neutron scattering is an ideal probe for investigating these aspects, but has thus far not been employed, in main due to the large background generated by spin incoherent scattering from $^1$H and $^{51}$V. We minimised this problem by replacing most of the $^1$H ($\sim98\%$) by $^2$D, which was achieved by performing the synthesis \cite{Hiroi2001} using deuterated reagents in an atmosphere of N$_2$, followed by annealing the product several times in D$_2$O at $95^\circ$C. Phase purity of the resulting sample was verified by powder XRD. The concentration of paramagnetic defects, proportional to the magnitude of the Curie tail \cite{Bert2004}, was estimated to be to be $<1\%$.

\begin{figure}[th]
\includegraphics[width=\linewidth]{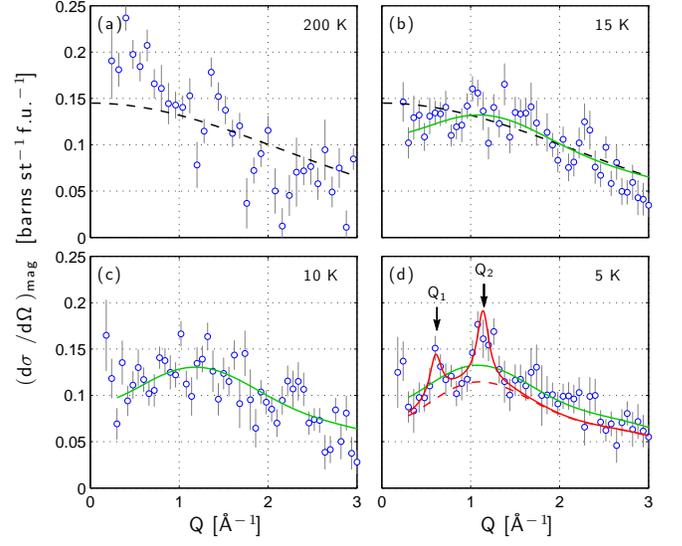}
\caption{(d$\sigma$/d$\Omega$)$_{\mathsf{mag}}$ for volborthite at $T=200,15,10,5$~K. (a) (d$\sigma$/d$\Omega$)$_{\mathsf{mag}}$ mimics $f(Q)^2$ (dashed black line) for Cu$^{2+}$. (b,c) Spectral weight shifts into a broad feature around $1.1-1.4$~\AA$^{-1}$. Green line is fit to Eq.\ (1). (d) Fit to (d$\sigma$/d$\Omega$)$_{\mathsf{mag}}$ at $5$~K as described in the text (red). The green line indicates the fit at $10$~K, showing the shift of spectral weight into the two sharp features at $Q_1$ and $Q_2$.}
\label{fig:d7tdep}
\end{figure}

Polarised diffuse neutron scattering was performed on D7 at ILL using $35.7$~g of sample and incident energy $E_i=8.5$~meV. Three orthogonal neutron polarisations ($x$, $y$, and $z$) and their corresponding spin flip and non-spin flip cross sections were analysed, which allowed for isolation of the magnetic scattering cross section, (d$\sigma$/d$\Omega$)$_{\mathsf{mag}}$ \cite{Stewart2009}. As no energy analysis was used, the observed scattering was effectively integrated up to $\omega=8.2$~meV, thus approximating the instantaneous structure factor, $S(Q)$. Spectra were measured at $T=200$~K, $15$~K, $10$~K, and $5$~K (Fig.\ \ref{fig:d7tdep}).

At $200$~K$\sim 2J$, the $Q$-dependence of (d$\sigma$/d$\Omega$)$_{\mathsf{mag}}$ approximately follows the Cu$^{2+}$ form factor, $|f(Q)|^2$, as anticipated for a paramagnet. Reducing the temperature to $15$~K, broad diffuse scattering develops around $Q=1.1-1.4$~\AA$^{-1}$. The $Q$-dependence is consistent with a buildup of nearest neighbour pair correlations, described by the powder averaged structure factor (d$\sigma$/d$\Omega$)$_{\mathsf{mag}}$:
\begin{equation}
\left(\frac{d \sigma}{d \Omega}\right)_{\mathsf{mag}} \!\!\!\!\!\!\!= \frac{2}{3}  \left( \frac{\gamma_n r_0}{2} \mu \right)^2 |f(Q)|^2  \left(1 + Z_1 \left<\mathbf{S}_0 \cdot \mathbf{S}_1 \right> \frac{\sin{Qr}}{Qr} \right)
\end{equation}
\noindent where the second term in the parenthesis reflects the average correlation $\left<\mathbf{S}_0 \cdot \mathbf{S}_1 \right>=-0.25(5)$ between a unit spin and its $Z_1$ nearest neighbours at a distance $r_{\mathsf{Cu-Cu}} \sim 3$~\AA. The total scattering was found to be $0.99(8)\mu_B^2$ per Cu, corresponding to 33\% of the full $S(S+1)$.The fact that correlations are weak and confined to only nearest neighbours even at $T/J_{avg} \sim 0.2$ are both indicators of strong frustration in volborthite.

As $T$ is further decreased to first $10$~K and then $5$~K, the broad diffuse scattering persists, but some ($14 \%$) of the spectral weight is shifted into two sharper (though not resolution limited) peaks at $Q_1=0.65(3)$~\AA$^{-1}$ and $Q_2=1.15(5)$~\AA$^{-1}$. The corresponding correlation length, $\xi=24(8)$~\AA$\sim 8r_{\mathsf{Cu-Cu}}$, was extracted by fitting the diffuse scattering to $(1)$ and the two sharper features to Lorentzians (Fig. \ref{fig:d7tdep}(d)).
\begin{figure*}
\includegraphics[width=0.95\linewidth]{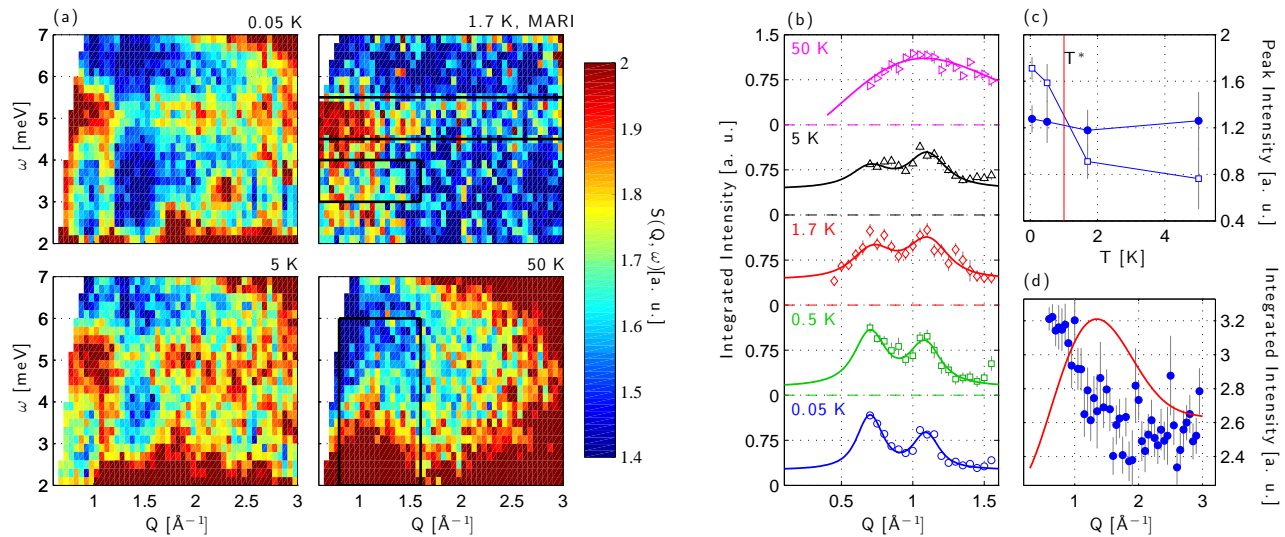}
\caption{(a) $S(Q,\omega)$ from IN4 at $T=0.05,~5,~50$~K, and MARI at $1.7$~K. The integration ranges used for the cuts in panels (b) and (d) are shown as black rectangles. (b) $Q$-dependence of intensity over energy ranges  and 3-4~meV for 0.05-5~K (fit by two resolution convoluted Lorentzians) and 1-6~meV for 50~K (fit by Eq.\ 1). (c) Temperature variation of intensities of the $Q_1$ (open circles) and $Q_2$ (closed circles) modes. $T^\ast=1$~K is indicated by a red vertical line. (d) $Q$-dependence of the flat mode integrated over 4.5-5.5~meV. $S(Q)$ for a singlet-triplet excitation is shown in red. }
\label{fig:in4}
\end{figure*}
The dynamical structure factor, $S(Q,\omega)$, was investigated by inelastic neutron scattering on IN4 at ILL. Spectra were collected using incident energy $E_i=17.2$~meV, giving elastic $Q$ range $0.65$~\AA$^{-1}-4.95$~\AA$^{-1}$. Further experiments were carried out on MARI at ISIS ($E_i=15$~meV, $0.45$~\AA$^{-1}-4.95$~\AA$^{-1}$ at $\omega = 0$~meV). Data are summarised in Fig. \ref{fig:in4}.

At $50$~K$\sim J/2$, $S(Q,\omega)$ shows only a broad response centered at $Q=1.1$~\AA$^{-1}$ and extending to $5$~meV, consistent with fluctuations in a short range correlated system. The $Q$-dependence extracted by integrating over the range $1-6$~meV is indicated in the top panel of Fig. \ref{fig:in4}(b). Its form is similar to $S(Q)$ at $10$~K and $15$~K, and can also be fitted by Eq. 1 using $r\sim 3.5(2)$~\AA$\sim r_{\mathsf{Cu-Cu}}$. Acoustic phonons are observed dispersing from nuclear Bragg positions at $Q>2$~\AA$^{-1}$ and intense phonon scattering is found above $7$~meV, making extraction of the magnetic signal at these energies difficult. Cooling to $5$~K, the low $Q$, low $\omega$ intensity has largely moved into two features: an intense broad flat band, centered at $\omega_f=5.0(2)$~meV, and a nearly vertical bar of scattering at $Q=1.08(2)$~\AA$^{-1}$, which coincides with the $Q_2$ peak in $S(Q)$. Both of these features sharpen as temperature is reduced towards $0.05$~K, with a second bar of scattering at $Q=0.68(4)$~\AA$^{-1}\sim Q_1$ growing below $1.7$~K. $Q$-cuts through the $Q_1$ and $Q_2$ modes are shown in the lower 4 panels of  Fig. \ref{fig:in4}(b). While the peaks narrow somewhat with decreasing $T$, a more dramatic change is observed in their respective intensities, $I(Q_1)$ and $I(Q_2)$. On cooling, $I(Q_2)$ remains constant, while $I(Q_1)$ increases to a final ratio $I(Q_1)/I(Q_2)=1.6$ at $0.05$~K. The buildup of dynamical correlations at $Q_1$ position thus coincides with the transition at $T^\ast$ observed in NMR and $\mu$SR. 

The lineshape and ampliude of the flat mode, on the other hand, show little temperature dependence, with only slight narrowing to become resolution limited between $5$~K and $1.7$~K. Such narrow flat modes are often associated with two level excitations, \emph{eg.} between a singlet and triplet. Indeed, such excitations are expected in the site diluted QKHAF \cite{Dommange2003,Singh2010}. While the predominantly nearest neighbour correlations observed in $S(Q)$ are consistent with such a state, the $Q$-dependence of the flat mode does not match the singlet-triplet $S(Q)$ \cite{Singh2010,Vries2009} [Fig. 3(d)]. Another possible explanation for the flat band is thus that it is associated with the short range order observed at the $Q_1$ and $Q_2$ positions. This leads to two scenarios: i.) That the ground state possesses a degeneracy which results in a flat band, as is the case in the ground state of the classical kagome systems \cite{Coomer2006a} or ii.) That the flat band arises as a consequence of powder averaging at the zone boundary of a spin wave dispersion.

In the pure QKHAF, no short range order is expected, contrary to observations. To understand the low $T$ short range ordered state, it is therefore necessary to look theoretically further afield. From structural considerations outlined in the introduction, the Hamiltonian of volborthite can include three exchange terms ($J_1,~J_1',~$and $J_2$) as well as a Dzyaloshinskii-Moriya (DM) interaction, for which the dominant component is considered to be $D_z$. The subsets of this model which have been treated semi-classically or quantum mechanically are: $J_1=J_1'> 0,J_2=0,D_z \neq 0$, the isotropic kagome DM model (IKDM) \cite{Cepas2008,Rousochatzakis2009,Messio2010} and $J_1 \neq J_1' > 0,J_2=0,D_z = 0$, the anisotropic kagome model (AK) \cite{Yavorskii2007,Wang2007}. In addition, recent L(S)DA+U calculations have suggested a model where $-J_1' / J_1 \sim 1.2-2$, $-J_2/J_1=1.1-1.6$, and $J_1'=8.6$~meV. This model, which we call the coupled chain model (CC), has only been treated classically so far \cite{Janson2010}. The $5$ possible ordered states which emerge from these models are: $\mathbf{q}=0$ order (IKDM for $D_z>0.1$, AK for $1/2<J_1'/J_1<1$), $\mathbf{q}=\sqrt{3}\times\sqrt{3}$ (AK for $1<J_1/J_1'<1.3$), chirality stripe (AK for $J_1/J_1'>1.3$), ferrimagnetic (AK for $J_1'/J_1<1/2$, CC for $J_2 < |J_1|/4 + J_1'/8$), and spiral (CC for $J_2 > |J_1|/4 + J_1'/8$). The position of Bragg peaks for these structures are compared with our experimental $Q_1$ and $Q_2$ in figure $4$(a). None of the proposed structures yield strong scattering at $Q_1$.
%\begin{align}
%\mathcal{H}=J_1\sum_{\langle i,j \rangle}&{\mathbf{S}_i\cdot \mathbf{S}_j}+J_1'\sum_{\langle i,j \rangle}{\mathbf{S}_i\cdot \mathbf{S}_j}... \nonumber \\
%&+J_2\sum_{\langle \langle i,j \rangle \rangle}{\mathbf{S}_i\cdot \mathbf{S}_j}+\sum_{\langle i,j \rangle}{[\mathbf{D_{ij}}\cdot\mathbf{S}_i\times\mathbf{S}_j]}
%\end{align}

Therefore, instead of attempting to describe the observed $S(Q,\omega)$ in terms of the microscopic model above, we take the phenomenological approach of constructing a generic dispersion emanating from antiferromagnetic zone centers at high symmetry positions on the circles defined by $Q_1$ and $Q_2$ (Fig.\ 4a). One example of such a dispersion among the scenarios considered assumes zone centers close to the $(10)$ and $(01)$ positions, yielding the dispersion:
\begin{equation}
\omega(\mathbf{q})= \sqrt{(2J^\mathsf{e}_\mathsf{a}+2J^\mathsf{e}_\mathsf{b})^2-[2J^\mathsf{e}_\mathsf{a}\cos(q_x a)+2J^\mathsf{e}_\mathsf{b}\cos(q_y b)]^2}
\end{equation}
where $J^{e}_{a,b}$ are effective exchanges, giving the amplitudes of the dispersion along $a$ and $b$. Then,
\begin{equation}
S(\mathbf{q},\omega)=|F(\mathbf{\tau})|^2 \frac{2-\cos(q_x a)-\cos(q_y b)}{\omega(\mathbf{q})}
\end{equation}
where $|F(\mathbf{\tau})|^2$ is the structure factor at the chosen positions, $1/\omega$ describes antiferromangetic spin wave intensity, and the numerator is a geometric term yielding zero intensity at ferromagnetic zone centers. To yield a smooth continuous function, $S(\mathbf{q},\omega)$ was interpolated between adjacent Brillouin zones. Finally, the spectrum was powder averaged and convoluted with the experimental resolution. The result of this procedure using $J^\mathsf{e}_\mathsf{a}=5.1$~meV and $J^\mathsf{e}_\mathsf{b}=15.3$~meV closely resembles the experimental data (Fig. 4(b)). One important conclusion from this analysis is that the steepness of the $Q_2$ mode implies that the $5$~meV band cannot be the the global zone boundary energy. Instead the flat band is found to be a saddle point, which requires sizeable anisotropy between $J^\mathsf{e}_\mathsf{a}$ and $J^\mathsf{e}_\mathsf{b}$. This result is consistent with LS(D)A+U calculations, which also suggest a significant anisotropy in exchange along $a$ and $b$.

\begin{figure}[t]
\includegraphics[width=0.95\linewidth]{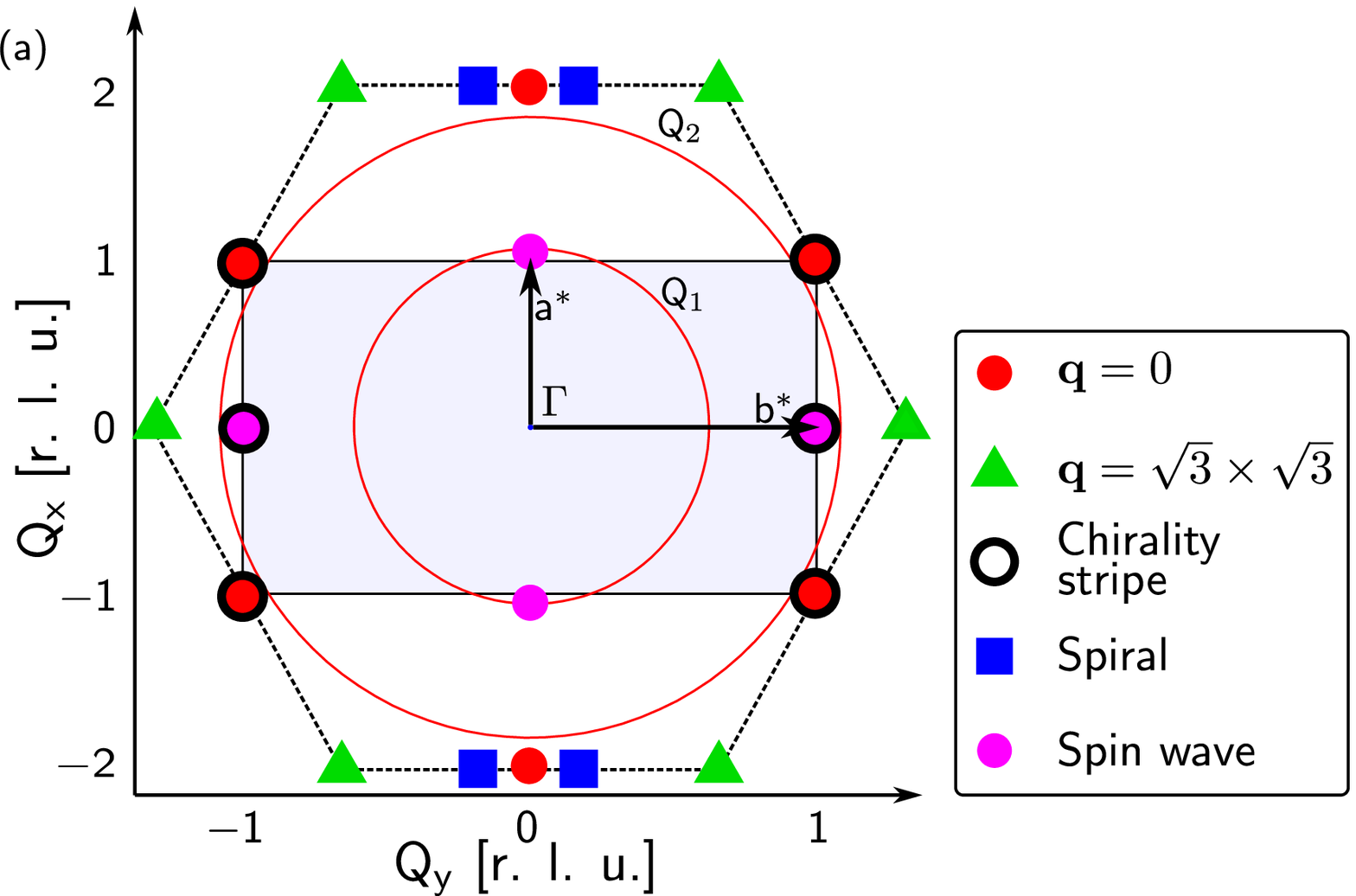}
\includegraphics[width=\linewidth]{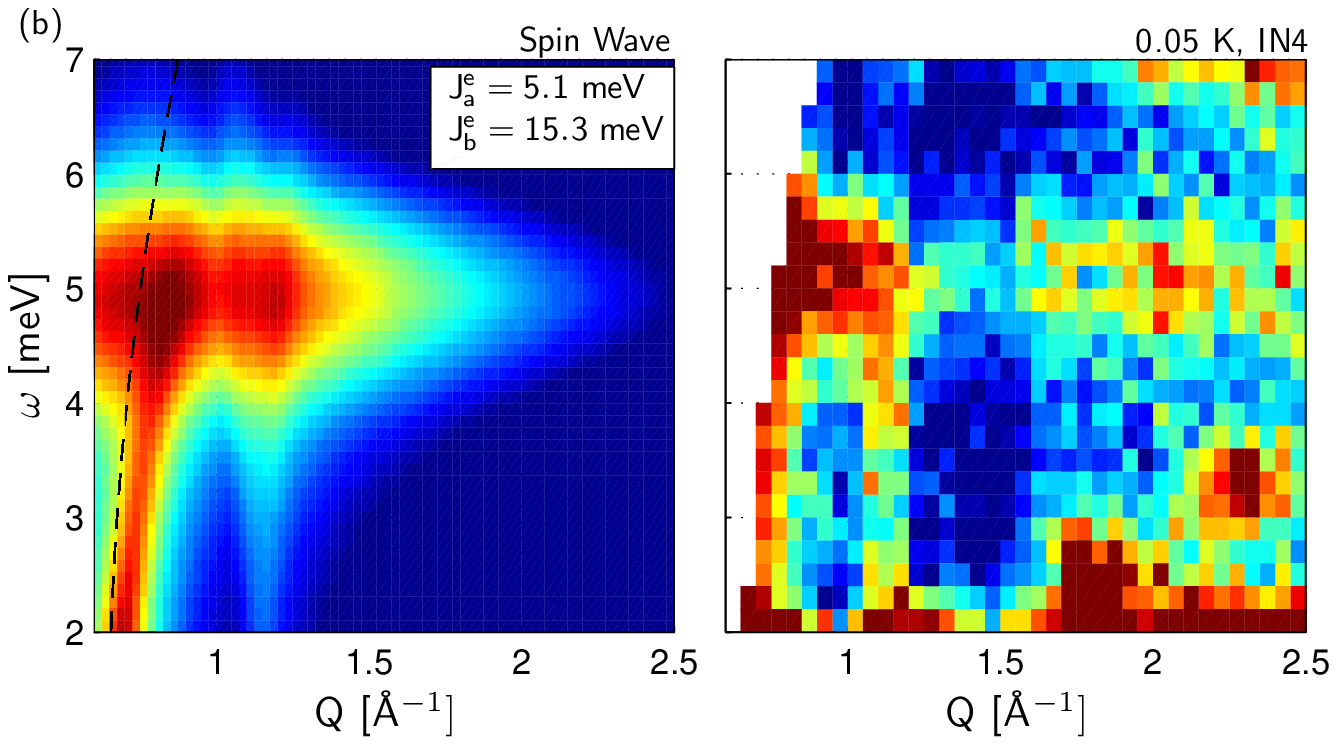}
\caption{(a) Reciprocal space of volborthite: the structural unit cell is indicated by the rectangle, and the extended Brillouin zone of the kagome lattice by the dotted hexagon. Symbols represent Bragg peaks of the orders listed in the text. (b) The experimental $S(Q,\omega)$ measured at $0.05$~K compared with the powder averaged $S(Q,\omega)$ derived from our empirical spin wave model. The dashed line in the left panel indicates the $(Q,\omega)$ window of the experiment.}
\label{fig:d7sf}
\end{figure}

In summary, we have reported polarised and inelastic neutron scattering results on the quasi-kagome $s=1/2$ antiferromagnet volborthite. These reveal three important features: i) Buildup of nearest neighbor pair correlations from $50$~K to $10$~K; ii) Short range order indicated by peaks at $Q_1=0.65(3)$~\AA$^{-1}$ and $Q_2=1.15(5)$~\AA$^{-1}$ in the diffuse and inelastic scattering below $5$~K; and iii) The excitation spectrum displays dispersive modes emanating from both $Q_1$ and $Q_2$ along with a flat mode at $\omega_f =5$~meV. The inelastic intensity at $Q_1$ becomes dominant below $1.7$~K, identifying the nature of the low-$T$ state reported from $^{51}$V NMR and $C_p$. Most models proposed for volborthite however inaccurately predict no scattering at $Q_1$, and could therefore be ruled out.  We provide an empirical dispersion model which reproduce the experimental observations for an anisotropic set of exchanges which are in rough agreement with recent LS(D)A+U calculations. A quantum treatment of this model could well yield a correct description of both the correlations and excitation spectrum that we have uncovered experimentally.

\begin{acknowledgments}
We thank Z. Hiroi, B. F\aa k, P. Mendels, F. Bert, and O. Janson for stimulating discussions, G. Ehlers and J. Gardner for assistance during initial experiments, and SNF for support.
\end{acknowledgments}

\bibliographystyle{apsrev}
\bibliography{volb}

\begin{thebibliography}{19}
\expandafter\ifx\csname natexlab\endcsname\relax\def\natexlab#1{#1}\fi
\expandafter\ifx\csname bibnamefont\endcsname\relax
  \def\bibnamefont#1{#1}\fi
\expandafter\ifx\csname bibfnamefont\endcsname\relax
  \def\bibfnamefont#1{#1}\fi
\expandafter\ifx\csname citenamefont\endcsname\relax
  \def\citenamefont#1{#1}\fi
\expandafter\ifx\csname url\endcsname\relax
  \def\url#1{\texttt{#1}}\fi
\expandafter\ifx\csname urlprefix\endcsname\relax\def\urlprefix{URL }\fi
\providecommand{\bibinfo}[2]{#2}
\providecommand{\eprint}[2][]{\url{#2}}

\bibitem[{\citenamefont{Shores et~al.}(2005)\citenamefont{Shores, Nytko,
  Bartlett, and Nocera}}]{Shores2005}
\bibinfo{author}{\bibfnamefont{M.}~\bibnamefont{Shores}},
  \bibinfo{author}{\bibfnamefont{E.}~\bibnamefont{Nytko}},
  \bibinfo{author}{\bibfnamefont{B.}~\bibnamefont{Bartlett}}, \bibnamefont{and}
  \bibinfo{author}{\bibfnamefont{D.}~\bibnamefont{Nocera}},
  \bibinfo{journal}{J. Am. Chem. Soc.} \textbf{\bibinfo{volume}{127}},
  \bibinfo{pages}{13462} (\bibinfo{year}{2005}).

\bibitem[{\citenamefont{Hiroi et~al.}(2001)\citenamefont{Hiroi, Hanawa,
  Kobayashi, Nohara, Takagi, Kato, and Takigawa}}]{Hiroi2001}
\bibinfo{author}{\bibfnamefont{Z.}~\bibnamefont{Hiroi}},
  \bibinfo{author}{\bibfnamefont{M.}~\bibnamefont{Hanawa}},
  \bibinfo{author}{\bibfnamefont{N.}~\bibnamefont{Kobayashi}},
  \bibinfo{author}{\bibfnamefont{M.}~\bibnamefont{Nohara}},
  \bibinfo{author}{\bibfnamefont{H.}~\bibnamefont{Takagi}},
  \bibinfo{author}{\bibfnamefont{Y.}~\bibnamefont{Kato}}, \bibnamefont{and}
  \bibinfo{author}{\bibfnamefont{M.}~\bibnamefont{Takigawa}},
  \bibinfo{journal}{J. Phys. Soc. Japan} \textbf{\bibinfo{volume}{70}},
  \bibinfo{pages}{3377} (\bibinfo{year}{2001}).

\bibitem[{\citenamefont{de~Vries et~al.}(2009)\citenamefont{de~Vries, Stewart,
  Deen, Piatek, Nilsen, R\o{}nnow, and Harrison}}]{Vries2009}
\bibinfo{author}{\bibfnamefont{M.~A.} \bibnamefont{de~Vries}},
  \bibinfo{author}{\bibfnamefont{J.~R.} \bibnamefont{Stewart}},
  \bibinfo{author}{\bibfnamefont{P.~P.} \bibnamefont{Deen}},
  \bibinfo{author}{\bibfnamefont{J.~O.} \bibnamefont{Piatek}},
  \bibinfo{author}{\bibfnamefont{G.~J.} \bibnamefont{Nilsen}},
  \bibinfo{author}{\bibfnamefont{H.~M.} \bibnamefont{R\o{}nnow}},
  \bibnamefont{and} \bibinfo{author}{\bibfnamefont{A.}~\bibnamefont{Harrison}},
  \bibinfo{journal}{Phys. Rev. Lett.} \textbf{\bibinfo{volume}{103}},
  \bibinfo{pages}{237201} (\bibinfo{year}{2009}).

\bibitem[{\citenamefont{Singh}(2010)}]{Singh2010}
\bibinfo{author}{\bibfnamefont{R.~R.~P.} \bibnamefont{Singh}},
  \bibinfo{journal}{Phys. Rev. Lett.} \textbf{\bibinfo{volume}{104}},
  \bibinfo{pages}{177203} (\bibinfo{year}{2010}).

\bibitem[{\citenamefont{Yavors'kii et~al.}(2007)\citenamefont{Yavors'kii, Apel,
  and Everts}}]{Yavorskii2007}
\bibinfo{author}{\bibfnamefont{T.}~\bibnamefont{Yavors'kii}},
  \bibinfo{author}{\bibfnamefont{W.}~\bibnamefont{Apel}}, \bibnamefont{and}
  \bibinfo{author}{\bibfnamefont{H.-U.} \bibnamefont{Everts}},
  \bibinfo{journal}{Phys. Rev. B} \textbf{\bibinfo{volume}{76}},
  \bibinfo{pages}{064430} (\bibinfo{year}{2007}).

\bibitem[{\citenamefont{Wang et~al.}(2007)\citenamefont{Wang, Vishwanath, and
  Kim}}]{Wang2007}
\bibinfo{author}{\bibfnamefont{F.}~\bibnamefont{Wang}},
  \bibinfo{author}{\bibfnamefont{A.}~\bibnamefont{Vishwanath}},
  \bibnamefont{and} \bibinfo{author}{\bibfnamefont{Y.~B.} \bibnamefont{Kim}},
  \bibinfo{journal}{Phys. Rev. B} \textbf{\bibinfo{volume}{76}},
  \bibinfo{pages}{094421} (\bibinfo{year}{2007}).

\bibitem[{\citenamefont{Cepas et~al.}(2008)\citenamefont{Cepas, Fong, Leung,
  and Lhuillier}}]{Cepas2008}
\bibinfo{author}{\bibfnamefont{O.}~\bibnamefont{Cepas}},
  \bibinfo{author}{\bibfnamefont{C.~M.} \bibnamefont{Fong}},
  \bibinfo{author}{\bibfnamefont{P.~W.} \bibnamefont{Leung}}, \bibnamefont{and}
  \bibinfo{author}{\bibfnamefont{C.}~\bibnamefont{Lhuillier}},
  \bibinfo{journal}{Phys. Rev. B} \textbf{\bibinfo{volume}{78}},
  \bibinfo{pages}{140405} (\bibinfo{year}{2008}).

\bibitem[{\citenamefont{Rousochatzakis
  et~al.}(2009)\citenamefont{Rousochatzakis, Manmana, Laeuchli, Normand, and
  Mila}}]{Rousochatzakis2009}
\bibinfo{author}{\bibfnamefont{I.}~\bibnamefont{Rousochatzakis}},
  \bibinfo{author}{\bibfnamefont{S.~R.} \bibnamefont{Manmana}},
  \bibinfo{author}{\bibfnamefont{A.~M.} \bibnamefont{Laeuchli}},
  \bibinfo{author}{\bibfnamefont{B.}~\bibnamefont{Normand}}, \bibnamefont{and}
  \bibinfo{author}{\bibfnamefont{F.}~\bibnamefont{Mila}},
  \bibinfo{journal}{Phys. Rev. B} \textbf{\bibinfo{volume}{79}},
  \bibinfo{pages}{214415} (\bibinfo{year}{2009}).

\bibitem[{\citenamefont{Messio et~al.}(2011)\citenamefont{Messio, Lhuillier,
  and Misguich}}]{Messio2011}
\bibinfo{author}{\bibfnamefont{L.}~\bibnamefont{Messio}},
  \bibinfo{author}{\bibfnamefont{C.}~\bibnamefont{Lhuillier}},
  \bibnamefont{and} \bibinfo{author}{\bibfnamefont{G.}~\bibnamefont{Misguich}},
  \bibinfo{journal}{arxiv:1101.1212}  (\bibinfo{year}{2011}).

\bibitem[{\citenamefont{Bert et~al.}(2005)\citenamefont{Bert, Bono, Mendels,
  Ladieu, Duc, Trombe, and Millet}}]{Bert2005}
\bibinfo{author}{\bibfnamefont{F.}~\bibnamefont{Bert}},
  \bibinfo{author}{\bibfnamefont{D.}~\bibnamefont{Bono}},
  \bibinfo{author}{\bibfnamefont{P.}~\bibnamefont{Mendels}},
  \bibinfo{author}{\bibfnamefont{F.}~\bibnamefont{Ladieu}},
  \bibinfo{author}{\bibfnamefont{F.}~\bibnamefont{Duc}},
  \bibinfo{author}{\bibfnamefont{J.}~\bibnamefont{Trombe}}, \bibnamefont{and}
  \bibinfo{author}{\bibfnamefont{P.}~\bibnamefont{Millet}},
  \bibinfo{journal}{Phys. Rev. Lett.} \textbf{\bibinfo{volume}{95}},
  \bibinfo{pages}{087203} (\bibinfo{year}{2005}).

\bibitem[{\citenamefont{Yoshida
  et~al.}(2009{\natexlab{a}})\citenamefont{Yoshida, Okamoto, Tayama,
  Sakakibara, Tokunaga, Matsuo, Narumi, Kindo, Yoshida, Takigawa
  et~al.}}]{Yoshida2009}
\bibinfo{author}{\bibfnamefont{H.}~\bibnamefont{Yoshida}},
  \bibinfo{author}{\bibfnamefont{Y.}~\bibnamefont{Okamoto}},
  \bibinfo{author}{\bibfnamefont{T.}~\bibnamefont{Tayama}},
  \bibinfo{author}{\bibfnamefont{T.}~\bibnamefont{Sakakibara}},
  \bibinfo{author}{\bibfnamefont{M.}~\bibnamefont{Tokunaga}},
  \bibinfo{author}{\bibfnamefont{A.}~\bibnamefont{Matsuo}},
  \bibinfo{author}{\bibfnamefont{Y.}~\bibnamefont{Narumi}},
  \bibinfo{author}{\bibfnamefont{K.}~\bibnamefont{Kindo}},
  \bibinfo{author}{\bibfnamefont{M.}~\bibnamefont{Yoshida}},
  \bibinfo{author}{\bibfnamefont{M.}~\bibnamefont{Takigawa}},
  \bibnamefont{et~al.}, \bibinfo{journal}{J. Phys. Soc. Japan}
  \textbf{\bibinfo{volume}{78}}, \bibinfo{pages}{043704}
  (\bibinfo{year}{2009}{\natexlab{a}}).

\bibitem[{\citenamefont{Yoshida
  et~al.}(2009{\natexlab{b}})\citenamefont{Yoshida, Takigawa, Yoshida, Okamoto,
  and Hiroi}}]{Yoshida2009a}
\bibinfo{author}{\bibfnamefont{M.}~\bibnamefont{Yoshida}},
  \bibinfo{author}{\bibfnamefont{M.}~\bibnamefont{Takigawa}},
  \bibinfo{author}{\bibfnamefont{H.}~\bibnamefont{Yoshida}},
  \bibinfo{author}{\bibfnamefont{Y.}~\bibnamefont{Okamoto}}, \bibnamefont{and}
  \bibinfo{author}{\bibfnamefont{Z.}~\bibnamefont{Hiroi}},
  \bibinfo{journal}{Phys. Rev. Lett.} \textbf{\bibinfo{volume}{103}},
  \bibinfo{pages}{077207} (\bibinfo{year}{2009}{\natexlab{b}}).

\bibitem[{\citenamefont{Yamashita et~al.}(2010)\citenamefont{Yamashita,
  Moriura, Nakazawa, Yoshida, Okamoto, and Hiroi}}]{Yamashita2010}
\bibinfo{author}{\bibfnamefont{S.}~\bibnamefont{Yamashita}},
  \bibinfo{author}{\bibfnamefont{T.}~\bibnamefont{Moriura}},
  \bibinfo{author}{\bibfnamefont{Y.}~\bibnamefont{Nakazawa}},
  \bibinfo{author}{\bibfnamefont{H.}~\bibnamefont{Yoshida}},
  \bibinfo{author}{\bibfnamefont{Y.}~\bibnamefont{Okamoto}}, \bibnamefont{and}
  \bibinfo{author}{\bibfnamefont{Z.}~\bibnamefont{Hiroi}}, \bibinfo{journal}{J.
  Phys. Soc. Japan} \textbf{\bibinfo{volume}{79}}, \bibinfo{pages}{083710}
  (\bibinfo{year}{2010}).

\bibitem[{\citenamefont{Bert et~al.}(2004)\citenamefont{Bert, Bono, Mendels,
  Trombe, Millet, Amato, Baines, and Hillier}}]{Bert2004}
\bibinfo{author}{\bibfnamefont{F.}~\bibnamefont{Bert}},
  \bibinfo{author}{\bibfnamefont{D.}~\bibnamefont{Bono}},
  \bibinfo{author}{\bibfnamefont{P.}~\bibnamefont{Mendels}},
  \bibinfo{author}{\bibfnamefont{J.}~\bibnamefont{Trombe}},
  \bibinfo{author}{\bibfnamefont{P.}~\bibnamefont{Millet}},
  \bibinfo{author}{\bibfnamefont{A.}~\bibnamefont{Amato}},
  \bibinfo{author}{\bibfnamefont{C.}~\bibnamefont{Baines}}, \bibnamefont{and}
  \bibinfo{author}{\bibfnamefont{A.}~\bibnamefont{Hillier}},
  \bibinfo{journal}{J. Phys.: Condens. Matter} \textbf{\bibinfo{volume}{16}},
  \bibinfo{pages}{S829} (\bibinfo{year}{2004}).

\bibitem[{\citenamefont{Stewart et~al.}(2009)\citenamefont{Stewart, Deen,
  Andersen, Schober, Barthelemy, Hillier, Murani, Hayes, and
  Lindenau}}]{Stewart2009}
\bibinfo{author}{\bibfnamefont{J.~R.} \bibnamefont{Stewart}},
  \bibinfo{author}{\bibfnamefont{P.~P.} \bibnamefont{Deen}},
  \bibinfo{author}{\bibfnamefont{K.~H.} \bibnamefont{Andersen}},
  \bibinfo{author}{\bibfnamefont{H.}~\bibnamefont{Schober}},
  \bibinfo{author}{\bibfnamefont{J.~F.} \bibnamefont{Barthelemy}},
  \bibinfo{author}{\bibfnamefont{J.~M.} \bibnamefont{Hillier}},
  \bibinfo{author}{\bibfnamefont{A.~P.} \bibnamefont{Murani}},
  \bibinfo{author}{\bibfnamefont{T.}~\bibnamefont{Hayes}}, \bibnamefont{and}
  \bibinfo{author}{\bibfnamefont{B.}~\bibnamefont{Lindenau}},
  \bibinfo{journal}{J. Appl. Crystallogr.} \textbf{\bibinfo{volume}{42}},
  \bibinfo{pages}{69} (\bibinfo{year}{2009}).

\bibitem[{\citenamefont{Dommange et~al.}(2003)\citenamefont{Dommange, Mambrini,
  Normand, and Mila}}]{Dommange2003}
\bibinfo{author}{\bibfnamefont{S.}~\bibnamefont{Dommange}},
  \bibinfo{author}{\bibfnamefont{M.}~\bibnamefont{Mambrini}},
  \bibinfo{author}{\bibfnamefont{B.}~\bibnamefont{Normand}}, \bibnamefont{and}
  \bibinfo{author}{\bibfnamefont{F.}~\bibnamefont{Mila}},
  \bibinfo{journal}{Phys. Rev. B} \textbf{\bibinfo{volume}{68}},
  \bibinfo{pages}{224416} (\bibinfo{year}{2003}).

\bibitem[{\citenamefont{Coomer et~al.}(2006)\citenamefont{Coomer, Harrison,
  Oakley, Kulda, Stewart, Stride, Fak, Taylor, and Visser}}]{Coomer2006a}
\bibinfo{author}{\bibfnamefont{F.~C.} \bibnamefont{Coomer}},
  \bibinfo{author}{\bibfnamefont{A.}~\bibnamefont{Harrison}},
  \bibinfo{author}{\bibfnamefont{G.~S.} \bibnamefont{Oakley}},
  \bibinfo{author}{\bibfnamefont{J.}~\bibnamefont{Kulda}},
  \bibinfo{author}{\bibfnamefont{J.~R.} \bibnamefont{Stewart}},
  \bibinfo{author}{\bibfnamefont{J.~A.} \bibnamefont{Stride}},
  \bibinfo{author}{\bibfnamefont{B.}~\bibnamefont{Fak}},
  \bibinfo{author}{\bibfnamefont{J.~W.} \bibnamefont{Taylor}},
  \bibnamefont{and} \bibinfo{author}{\bibfnamefont{D.}~\bibnamefont{Visser}},
  \bibinfo{journal}{J. Phys.: Condens. Matter} \textbf{\bibinfo{volume}{18}},
  \bibinfo{pages}{8847} (\bibinfo{year}{2006}).

\bibitem[{\citenamefont{Messio et~al.}(2010)\citenamefont{Messio, Cepas, and
  Lhuillier}}]{Messio2010}
\bibinfo{author}{\bibfnamefont{L.}~\bibnamefont{Messio}},
  \bibinfo{author}{\bibfnamefont{O.}~\bibnamefont{Cepas}}, \bibnamefont{and}
  \bibinfo{author}{\bibfnamefont{C.}~\bibnamefont{Lhuillier}},
  \bibinfo{journal}{Phys. Rev. B} \textbf{\bibinfo{volume}{81}},
  \bibinfo{pages}{064428} (\bibinfo{year}{2010}).

\bibitem[{\citenamefont{Janson et~al.}(2010)\citenamefont{Janson, Richter,
  Sindzingre, and Rosner}}]{Janson2010}
\bibinfo{author}{\bibfnamefont{O.}~\bibnamefont{Janson}},
  \bibinfo{author}{\bibfnamefont{J.}~\bibnamefont{Richter}},
  \bibinfo{author}{\bibfnamefont{P.}~\bibnamefont{Sindzingre}},
  \bibnamefont{and} \bibinfo{author}{\bibfnamefont{H.}~\bibnamefont{Rosner}},
  \bibinfo{journal}{Phys. Rev. B} \textbf{\bibinfo{volume}{82}},
  \bibinfo{pages}{104434} (\bibinfo{year}{2010}).

\end{thebibliography}

\end{document}